\documentclass[prl,amsmath,amssymb, twocolumn, showpacs]{revtex4}

\usepackage{dcolumn}

\usepackage{bm}

\usepackage[dvips]{graphicx}

\usepackage{epsfig}

\begin{document}

\title{Decoherence in a driven three-level system}

\author{A.\  R.\  P. Rau$^{*}$ and Weichang Zhao}
\affiliation{Department of Physics and Astronomy, Louisiana State University,
Baton Rouge, Louisiana 70803-4001}

\date{\today}

\begin{abstract}

Dissipation and decoherence, and the evolution from pure to mixed states in
quantum physics are handled through master equations for the density matrix. Master
equations such as the Lindblad equation preserve the trace of this matrix. Viewing them
as first-order time-dependent operator equations for the elements of the density matrix,
a unitary integration procedure can be adapted to solve for these elements. A simple
model for decoherence preserves the hermiticity of the density matrix. A single, classical
Riccati equation is the only one requiring numerical handling to obtain a full solution of the quantum evolution. The procedure is general, valid for any number of levels, but is illustrated here for a three-level system with two driving fields. For various choices of the initial state, we study the evolution of the system as a function of the amplitudes, relative frequencies and phases of the driven fields, and of the strength of the decoherence. The monotonic growth of the entropy is followed as the system evolves from a pure to a mixed state.  An example is provided by the $n=3$ states of the hydrogen atom in a time-dependent electric field, such degenerate manifolds affording an analytical solution.
\end{abstract}


\maketitle

\section{Unitary integration procedure for master equations}

Master equations, such as the Lindblad equation \cite{ref1}, can describe dissipation and decoherence in quantum systems. In recent work \cite{ref2}, one of us adapted a ``unitary
integration" procedure \cite{ref3, ref4} for solving such equations while preserving desirable
properties such as the hermiticity of the density matrix even in the presence of dissipation
and decoherence. This permits keeping track of quantities
such as the entropy while the system evolves from a possibly initial pure state to a final mixed one.
The two-state illustration given in that initial work is extended now to a three-level system through suitable combinations of density matrix elements to preserve the hermiticity of the operators involved.  
 
Consider the master equation for the density matrix $\rho $
called the Liouville-von Neumann-Lindblad equation \cite{ref1, ref2},

\begin{eqnarray}
i\dot{\rho} & = & [H,\rho ]+
\frac{1}{2}i\!\sum_{k}\left( [L_k\rho,L_k^{\dagger }]+
[L_k,\rho L_k^{\dagger }]\right)  \nonumber \\
& = & [H,\rho ]-\frac{1}{2}i\!\sum_{k}\left( L_k^{\dagger }L_k\rho +\rho
L_k^{\dagger }L_k-2L_k\rho L_k^{\dagger }\right)\!, \label{eqn1}
\end{eqnarray}

\noindent where an over-dot denotes differentiation with respect to time and $\hbar$
 has been set equal to unity. $H$ is a Hermitian Hamiltonian while the $L_k$ are operators
 in the system through which dissipation and decoherence are introduced. Even though
 this can result in non-unitary evolution, the form of the equation preserves Tr($\rho$) and
 positivity of probabilities. A more mathematical discussion of such ``super-operators" and
 ``dynamical semigroups" is given in \cite{ref5}.
 
 A commonly used form of $H$ is
 
 \begin{equation}
 H(t)=\epsilon(t)A_z+2J(t)A_x,
 \label{eqn2}
 \end{equation}
 with
 
 \begin{eqnarray}
A_x=\left( 
\begin{array}{ccc}
0 & 0 & 0 \\ 
0 & 0 & 1 \\ 
0 & 1 & 0
\end{array}
\right)&,& A_y=\left( 
\begin{array}{ccc}
0 & 0 & -i \\ 
0 & 0 & 0 \\ 
i & 0 & 0
\end{array}
\right),\, \nonumber \\
A_z &=& \left( 
\begin{array}{ccc}
0 & 1 & 0 \\ 
1 & 0 & 0 \\ 
0 & 0 & 0
\end{array}
\right). 
\label{eqn3}
\end{eqnarray}
 
\noindent The couplings indicated in Eq.~(\ref{eqn2}) between states 1 and 2 and between 2 and
3 of a three-state system are referred to as $\Lambda$ and $V$ depending on the relative
energy positions of the three states, whether 2 lies above or below, respectively, relative to
levels 1 and 3. The three operators in Eq.~(\ref{eqn3}) close under commutation according to
the standard relations satisfied by angular momentum algebra: $[A_x,A_y]=iA_z$, and cyclic. Hioe and Eberly \cite{ref6} considered such a Hamiltonian for the Liouville version of Eq.~(\ref{eqn1}), that is, without the dissipative term, along with solutions for certain forms of $\epsilon$ and $J$. Population trapping and dispersion was also considered in \cite{ref7} with a similar Hamiltonian, and \cite{ref8} generalized to $n$-level systems. Our work presented here may be regarded as extending such studies to include also dissipation and decoherence. 

In general, with each of the three states having distinct energies $E_1,E_2,E_3$, and the driving fields having finite detunings from resonance, entries along the diagonal of $H$ in Eq.~(\ref{eqn2}) complete the Hamiltonian for such systems. Full 
treatment according to our formalism below then requires all the elements of the SU(3) algebra, 
namely five more linearly independent $3\times 3$ matrices to supplement those in Eq.~(\ref{eqn3}). We expect to return to this later but, in this paper, we restrict ourselves to the degenerate case of equal eigenvalues
in which case the above three matrices suffice and the calculations reduce to solving a single
equation just as in the two-state system considered in \cite{ref2}. Applications include three 
identical coupled pendula with nearest neighbor time-dependent couplings, driven systems on resonance, and the degenerate states of the $n=3$ manifold of hydrogen driven by time-dependent electric fields.

Dissipation and decoherence are introduced through the $L_k$ matrices in Eq.~(\ref{eqn1}). 
Here again, as shown in \cite{ref2}, a choice of all eight linearly independent matrices affords
a simplification because of a sum rule that inserts the decoherence as a unit operator in such an eight-dimensional space. In this procedure, Eq.~(\ref{eqn1}) is recast into a set 
of eight equations for the elements of the density matrix (recall that the trace remains invariant).
An appropriate linear combination of the elements such that the operators in Eq.~(\ref{eqn3}) 
map onto three Hermitian $8\times 8$ matrices obeying the same angular momentum commutators is given by the choice

\begin{eqnarray}
\eta(t)\!\! &=& \!\!(\rho_{11}\!-\!\rho_{33}, \!\frac{1}{\sqrt 3}(\rho_{11}\!+\!\rho_{33}\!-\!2\rho_{22}),
\rho_{12}\!+\!\rho_{21}, \rho_{21} \nonumber \\
 & &\!\! -\!\rho_{12}, \rho_{13}\!+\!\rho_{31}, \rho_{31}\!-\!\rho_{13}, 
\rho_{23}\!+\!\rho_{32}, \rho_{32}\!-\!\rho_{23}).
\label{eqn4}
\end{eqnarray}

Our choice differs only slightly from that in \cite{ref6, ref8}where this set is called a ``coherence" vector. The resulting equation for $\eta(t)$ takes the form

\begin{equation}
i\dot{\eta}(t)=\mathcal{L}(t) \eta(t),
\label{eqn5}
\end{equation}
with

\begin{equation}
\mathcal{L}(t)=-i\Gamma \mathcal{I} +\epsilon(t) B_z +2J(t) B_x,
\label{eqn6}
\end{equation}
where $\Gamma$ indexes the strength of the decoherence. The matrices $B$ take the form

\begin{eqnarray*}
B_x &=& \left( 
\begin{array}{cccccccc}
0 & 0 & 0 & 0 & 0 & 0 & 0 & 1 \\ 
0 & 0 & 0 & 0 & 0 & 0 & 0 & -\sqrt 3 \\ 
0 & 0 & 0 & 0 & 0 & 1 & 0 & 0 \\
0 & 0 & 0 & 0 & 1 & 0 & 0 & 0 \\
0 & 0 & 0 & 1 & 0 & 0 & 0 & 0 \\  
0 & 0 & 1 & 0 & 0 & 0 & 0 & 0 \\
0 & 0 & 0 & 0 & 0 & 0 & 0 & 0 \\
1 & -\sqrt 3 & 0 & 0 & 0 & 0 & 0 & 0 
\end{array}
\right), 
\end{eqnarray*}

\begin{eqnarray}
B_y &=& \left( 
\begin{array}{cccccccc}
0 & 0 & 0 & 0 & -2i & 0 & 0 & 0 \\ 
0 & 0 & 0 & 0 & 0 & 0 & 0 & 0 \\ 
0 & 0 & 0 & 0 & 0 & 0 & -i & 0 \\
0 & 0 & 0 & 0 & 0 & 0 & 0 & i \\
2i & 0 & 0 & 0 & 0 & 0 & 0 & 0 \\  
0 & 0 & 0 & 0 & 0 & 0 & 0 & 0 \\
0 & 0 & i & 0 & 0 & 0 & 0 & 0 \\
0 & 0 & 0 & -i & 0 & 0 & 0 & 0
\end{array}  
\right), \nonumber \\
B_z &=& \left( 
\begin{array}{cccccccc}
0 & 0 & 0 & 1 & 0 & 0 & 0 & 0 \\ 
0 & 0 & 0 & \sqrt 3 & 0 & 0 & 0 & 0 \\ 
0 & 0 & 0 & 0 & 0 & 0 & 0 & 0 \\
1 & \sqrt 3 & 0 & 0 & 0 & 0 & 0 & 0 \\
0 & 0 & 0 & 0 & 0 & 0 & 0 & -1 \\  
0 & 0 & 0 & 0 & 0 & 0 & -1 & 0 \\
0 & 0 & 0 & 0 & 0 & -1 & 0 & 0 \\
0 & 0 & 0 & 0 & -1 & 0 & 0 & 0
\end{array}  
\right). 
\label{eqn7}
\end{eqnarray}
           
As per the unitary integration procedure \cite{ref2,ref3}, the solution of  Eq.~(\ref{eqn6})
is written as a product of exponentials

\begin{eqnarray}
\eta (t)& = &\exp [-\Gamma t] \exp [-i\mu _{+}(t)B_{+}] \nonumber \\
&& \times 
\exp [-i\mu_{-}(t)B_{-}] 
\exp [-i\mu (t)B_z]\eta (0),  \label{eqn8}
\end{eqnarray}

\noindent with $B_{\pm }\equiv B_x\pm iB_y$. Because our procedure depends only on the
commutation relations which remain as in \cite{ref2}, the classical functions $\mu$ in the exponents satisfy the same equations as before,

\begin{subequations}
\begin{eqnarray}
\dot{\mu}_{+}-i\epsilon (t)\mu _{+}-J(t)(1+\mu _{+}^2)
&=&0 ,  \label{subeq1} \\
\dot{\mu}=2iJ(t)\mu _{+}-\epsilon (t) , &&  \label{subeq2} \\
\dot{\mu}_{-}-i\dot{\mu}\mu _{-}=J(t) ,\; && \mu _i(0)=0.  \label{subeq3}
\end{eqnarray}
\end{subequations}

\noindent The first of these equations, involving $\mu _{+}(t)$ alone
in Riccati form, is the only non-trivial member of this set. Once solved, $\mu_{-}$ and $\mu$ are obtained through simple integration of the remaining two equations. For given $\epsilon(t)$ and $J(t)$, a mathematica program solves the set of equations readily. Also, the subsequent algebra involved in evaluating the exponentials in Eq.~(\ref{eqn8}) and their product is easily carried out. Thereby, for any initial density matrix and its $\eta(0)$, we obtain $\eta(t)$ and thus $\rho(t)$ at any later time.

Since our model for decoherence introduces its effect through the single real factor which is the first term on the right-hand side of Eq.~(\ref{eqn8}), the density matrix remains Hermitian throughout. This is an advantage, permitting evaluation of its eigenvalues and calculation of quantities such as the entropy of the system. It is also clear that for any finite $\Gamma$ all elements in $\eta(t)$ in Eq.~(\ref{eqn4}) vanish asymptotically with $t$ so that all off-diagonal elements of the density matrix so vanish while all diagonal elements become equal. With the trace invariant and chosen to be unity, the density matrix evolves to that of the so-called chaotically mixed state, $\frac{1}{3} \mathcal{I}$. Correspondingly, the entropy reaches asymptotically the value $\ln 3 = 1.0986$. These are aspects of the general result valid for all $n$-level systems \cite{ref2}. We note again that other models of decoherence and dissipation through other choices for the operators $L_k$ in Eq.~(\ref{eqn1}) than the one we made will, in general, lead to a larger set of exponential factors in Eq.~(\ref{eqn8}), making for more complicated algebra therein and in the coupled set of equations in Eq.\ (9). However, inclusion of a term involving also $A_y$ in Eq.~(\ref{eqn2}), that is, a coupling also between levels 1 and 3, causes no additional difficulty since it does not enlarge the number of $A$ or $B$ matrices in our procedure.

\section{Two different driving fields between neighboring states}

We present results for three degenerate states, such as of three identical pendula, with different nearest neighbor couplings between 1-2 and 2-3, that is, with $\epsilon(t)$ and $J(t)$ differing in amplitude and frequency,

\begin{equation}
\epsilon(t)=A \cos \Omega t, \,\, J(t)=\frac{1}{2} B \cos (\omega t + \delta),
\label{eqn10}
\end{equation}
with $\delta$ a relative phase difference.
A representative sample of the density matrix upon starting with all population in the state 1 and all other elements zero is shown in Figs.\ 1-4. Note the appearance of a complicated frequency spectrum beyond just the two introduced driving frequencies. The analytically solvable problem presented in the next section provides an understanding of the origin of these other frequencies. As shown in Fig.\ 3, the entire population can be transferred from level 1 to level 3 over certain time intervals. 

\begin{figure}
\includegraphics[width=3in]{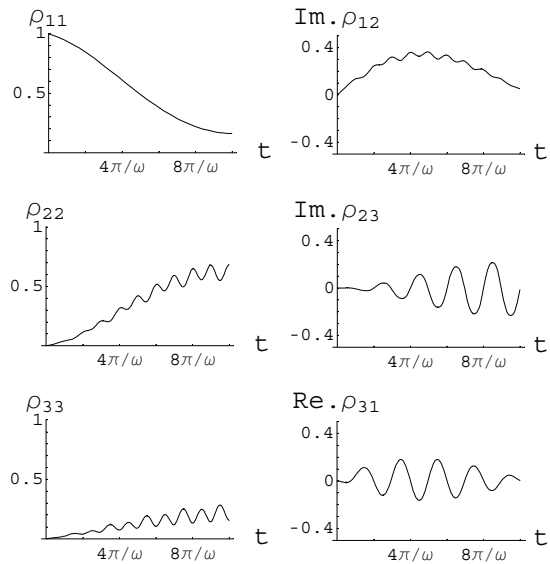}
\caption{Time evolution of the elements of the density matrix of a $n=3$ system driven by the fields in Eq.~(\ref{eqn10}) and Hamiltonian in Eq.~(\ref{eqn2}), with $\Omega =0, \omega =1, \delta =0, A =0.05, B= 0.5, \Gamma =0.02$. Right hand panels show the off-diagonal elements, two of which are imaginary and one real.}
\end{figure}

\begin{figure}
\includegraphics[width=3in]{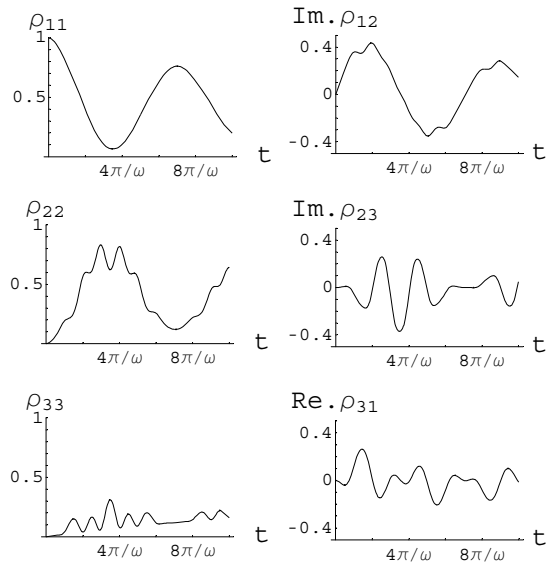}
\caption{same as in Fig.\ 1, except that $A =0.15$.}
\end{figure}

\begin{figure}
\includegraphics[width=3in]{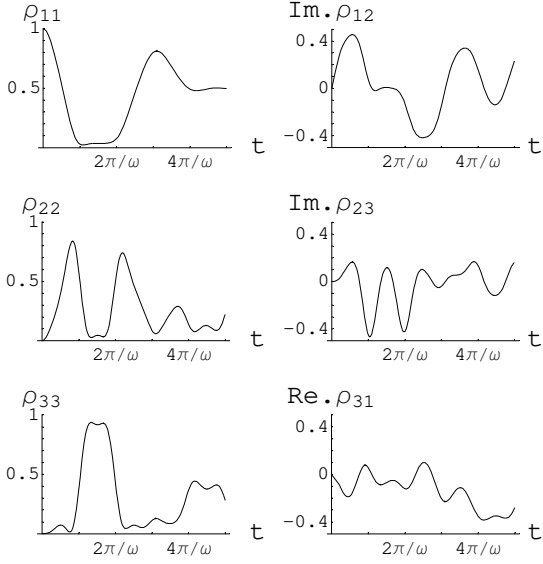}
\caption{same as in Fig.\ 1, except that $A =0.5, B =1, \Omega =0.1, \omega =1$.}
\end{figure}

\begin{figure}
\includegraphics[width=3in]{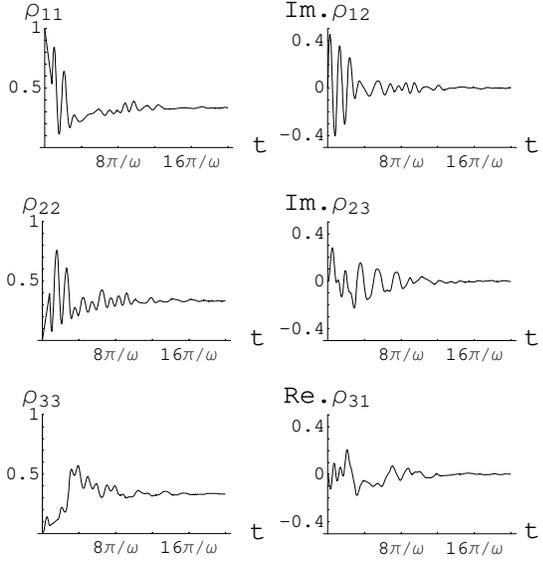}
\caption{same as in Fig.\ 3 except that $A =1, B =1/\sqrt{2}, \Gamma =0.08$ and longer times shown to illustrate asymptotic evolution.}
\end{figure}

Specializing to equal driving frequencies with a fixed amplitude ratio, results for various phase differences between the two fields are shown in Figs.\ 5-8. Clearly, the density matrix elements depend on the relative phase.

\begin{figure}
\includegraphics[width=3in]{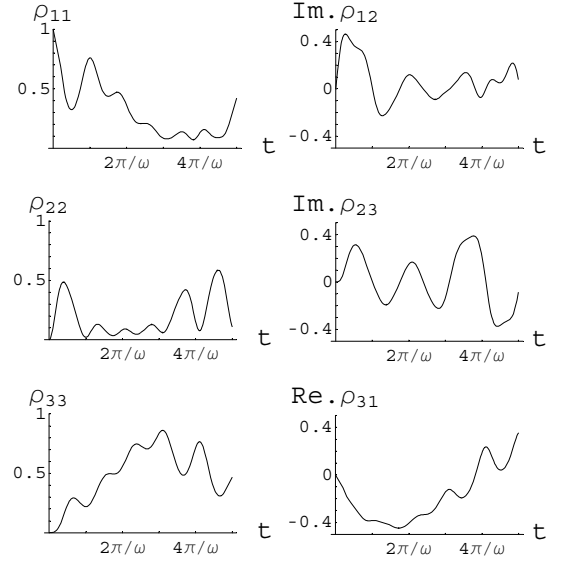}
\caption{Elements of the density matrix with driving fields of same frequency, $\Omega =\omega =1$, and amplitudes $A =1, B =1/\sqrt{2}$, and $\Gamma =0.02, \delta =-\pi/6$. }
\end{figure}

\begin{figure}
\includegraphics[width=3in]{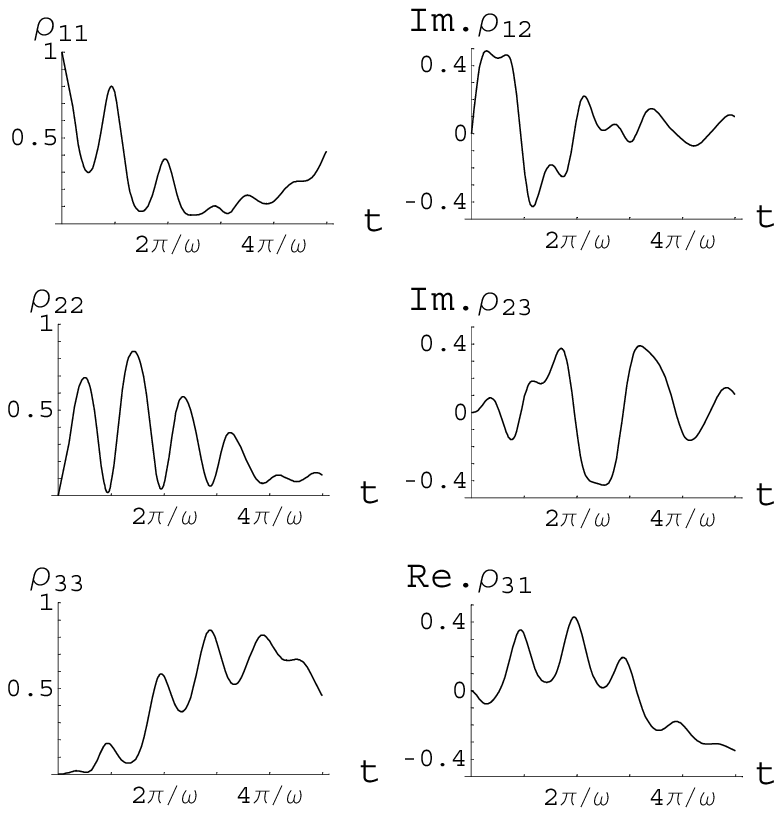}
\caption{same as in Fig.\ 5 except that $\delta =\pi/6$.}
\end{figure}

\begin{figure}
\includegraphics[width=3in]{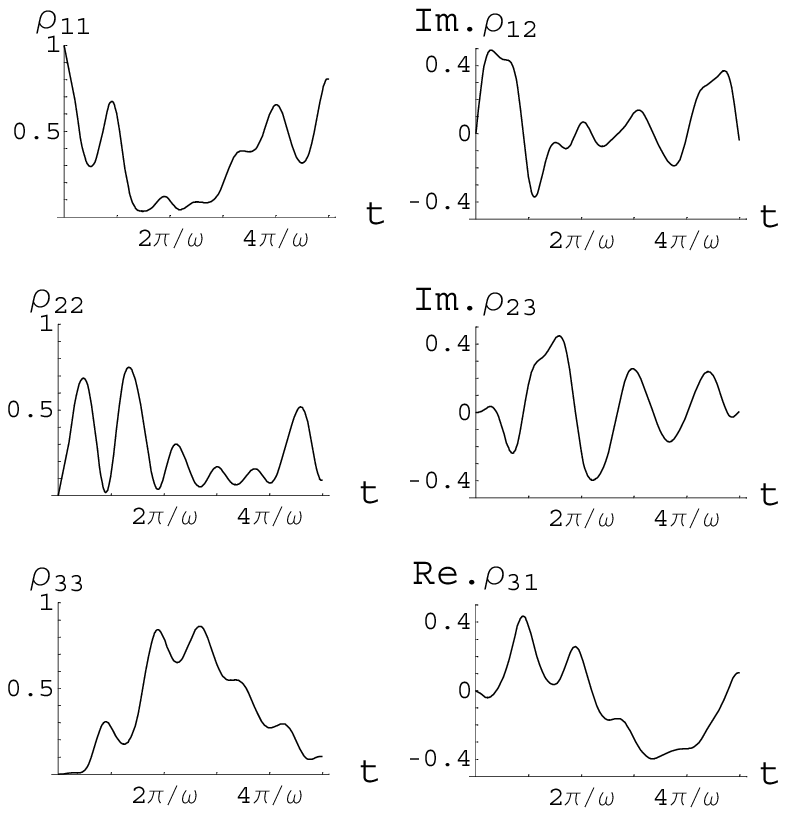}
\caption{same as in Fig.\ 5 except that $\delta =\pi/4$.}
\end{figure}

\begin{figure}
\includegraphics[width=3in]{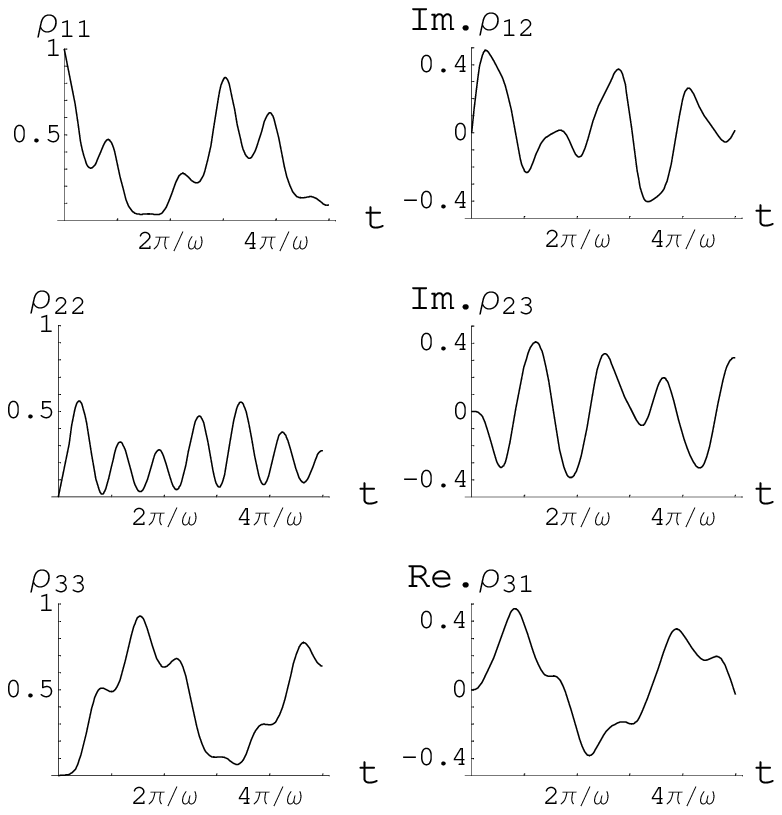}
\caption{same as in Fig.\ 5 except that $\delta =\pi/2$.}
\end{figure}

To contrast with a different initial state, Figs.\ 9-10 show results when all population is in the state 2 at $t=0$. 

\begin{figure}
\includegraphics[width=3in]{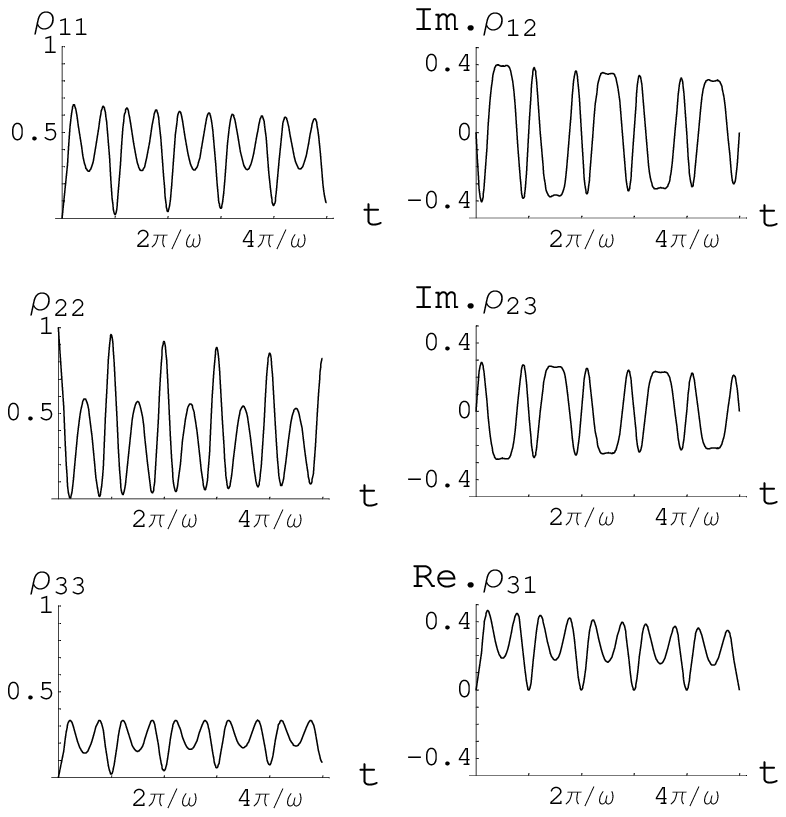}
\caption{Evolution of density matrix elements, starting with an initial non-zero value only for $\rho_{22} =1$. Contrast with Figs.\ 1-4. The parameters are $\Omega =\omega =1, A=2, B=\sqrt{2}, \Gamma =0.02, \delta =0$.}
\end{figure}

\begin{figure}
\includegraphics[width=3in]{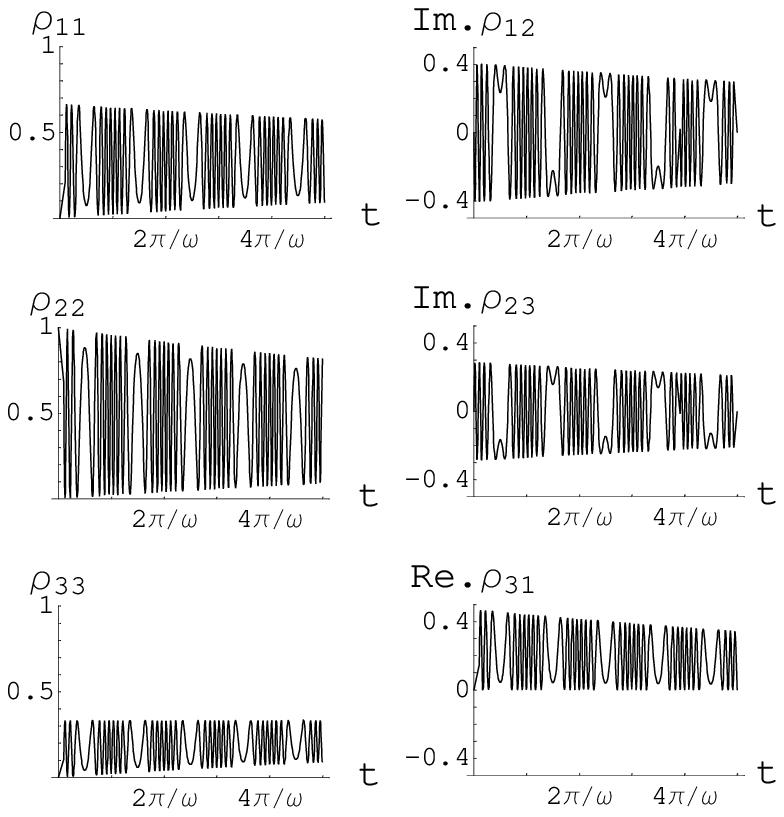}
\caption{same as in Fig.\ 9 except that $A=10, B=5\sqrt{2}$. Note the appearance of more rapid oscillations with the harder driving fields.}
\end{figure}

\noindent Fig.\ 11 presents the evolution of the entropy,
$S=-{\rm Tr}\rho \ln \rho$, showing a monotonic rise independent of amplitudes and phases and of the initial pure state. Indeed, the eigenvalues of the density matrix are $\frac{1}{3}(1-e^{-\Gamma t}), \frac{1}{3}(1-e^{-\Gamma t})$, and $\frac{1}{3}(1+2e^{-\Gamma t})$, from which the entropy easily follows.

\begin{figure}
\includegraphics[width=3in]{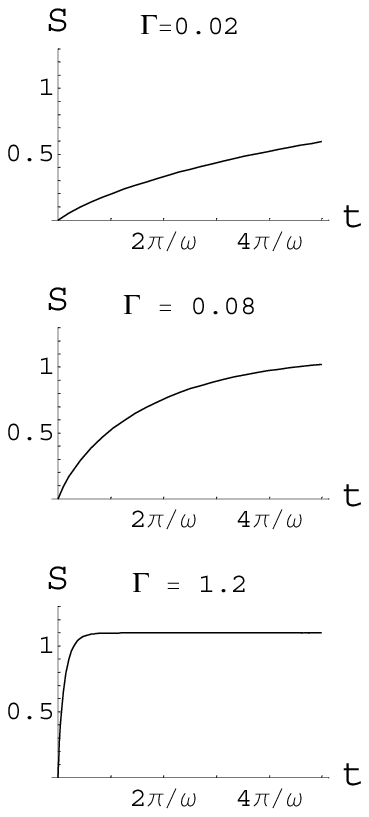}
\caption{Evolution of the entropy to accompany the results shown in previous figures. The rise is monotonic from $0$ to $\ln 3$, the rate of rise depending only on the value of $\Gamma$.}
\end{figure}

\section{$n=3$ states of the hydrogen atom in an oscillating electric field}

An example of  a three-state degenerate system is provided by the $n=3, m=0$ states of the hydrogen atom. An oscillating electric field such as that of incident radiation couples $3s-3p$ and $3p-3d$ states, the dipole matrix elements being in the ratio $\sqrt 2:1$. Our results in this paper apply to this situation with the two frequencies in Eq.~(\ref{eqn10}) equal and $A/B=\sqrt 2$. These were the choices made in Figs.\ 5-8. We present in Figs.\ 12-15 a sample of results for initial population in $3s$ for different amplitudes of the driving field.

\begin{figure}
\includegraphics[width=3in]{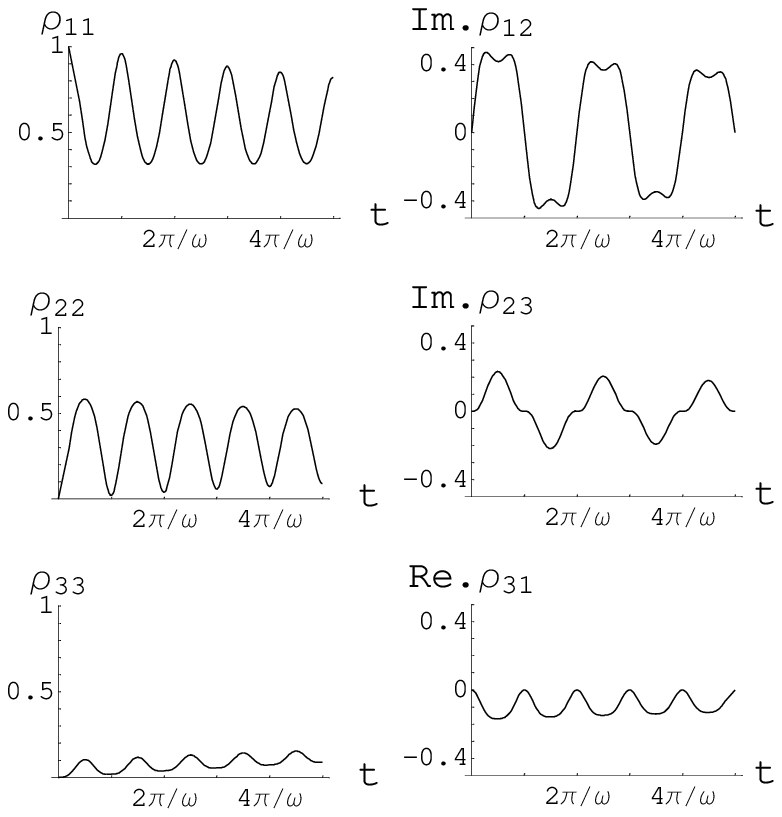}
\caption{Evolution of the density of states of the $n=3$ Stark field with initial population in the $3s$ state. The amplitude of the driving field is $A=1$ and $\Gamma =0.02$.}
\end{figure}

\begin{figure}
\includegraphics[width=3in]{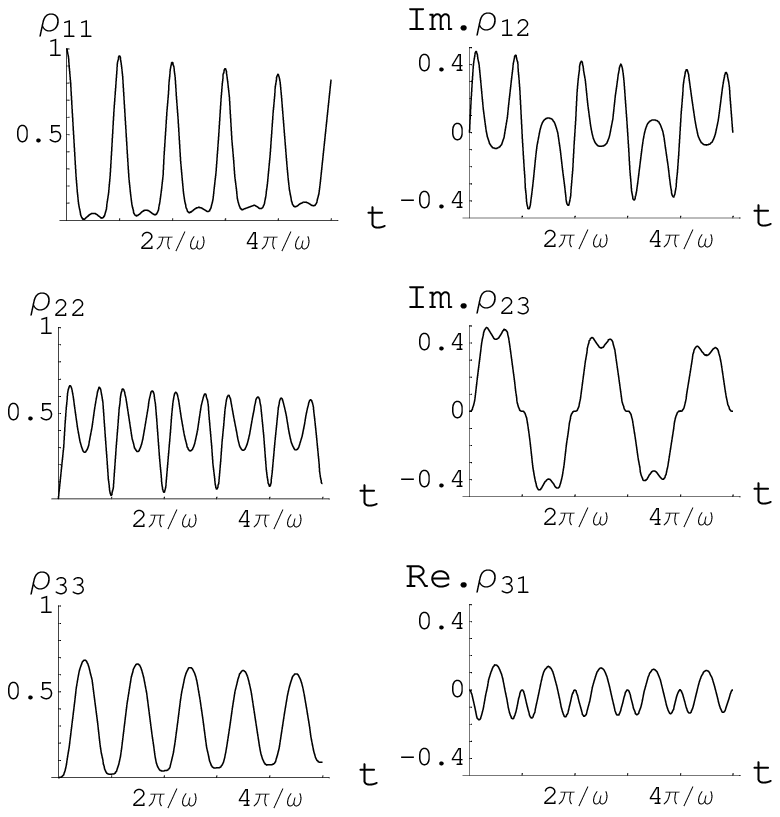}
\caption{same as in Fig.\ 12 except that $A=2$.}
\end{figure}

\begin{figure}
\includegraphics[width=3in]{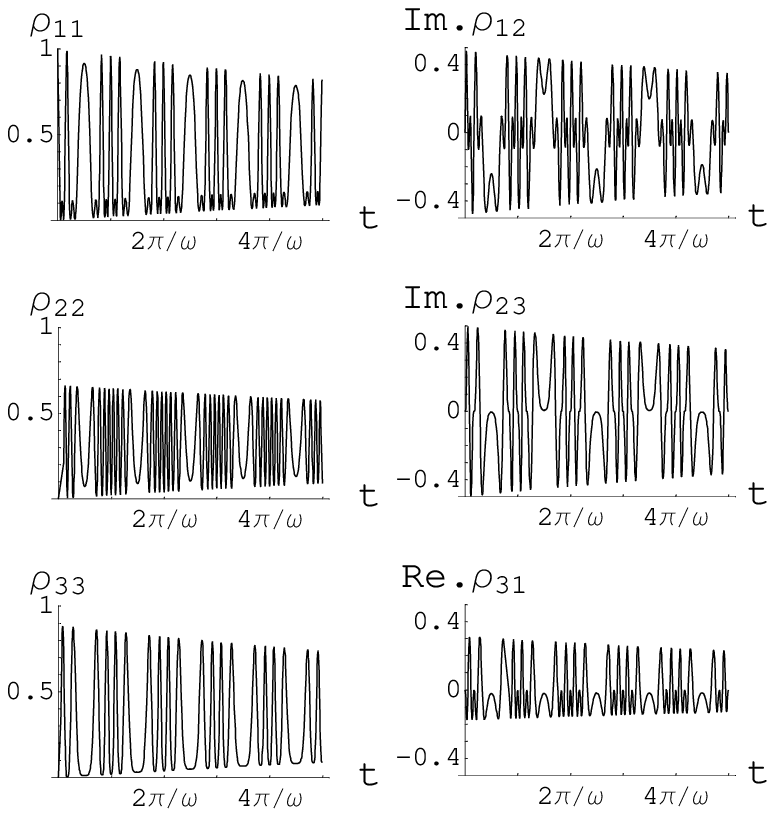}
\caption{same as in Fig.\ 12 except that $A=10$.}
\end{figure}

\begin{figure}
\includegraphics[width=3in]{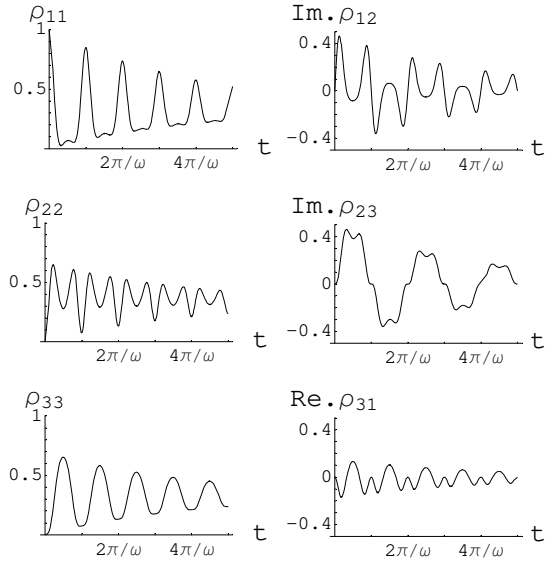}
\caption{same as in Fig.\ 13 except with a larger $\Gamma =0.08$.}
\end{figure}

This problem is, of course, exactly solvable in terms of the parabolic eigenstates of hydrogen. With $H(t)$ in Eq.~(\ref{eqn2}) containing a single time dependence, the resulting Schr\"{o}dinger equation,

\begin{equation}
i\left( 
\begin{array}{c}
\dot{s}(t) \\ 
\dot{p}(t) \\ 
\dot{d}(t)
\end{array}
\right) =\left( 
\begin{array}{ccc}
0 & -A & 0 \\ 
-A & 0 & -A/\sqrt 2 \\ 
0 & -A/\sqrt 2 & 0
\end{array}
\right) \left(
\begin{array}{c}
s(t) \\
p(t) \\
d(t)
\end{array}
\right)
\cos \omega t,
\label{eqn11}
\end{equation}
can be solved after diagonalizing the matrix of constant coefficients to obtain the parabolic eigenstates $\{\frac{1}{\sqrt 3} s \pm \frac {1}{\sqrt 2} p+\frac{1}{\sqrt 6} d, \frac{1}{\sqrt 3}(s-\sqrt{2} d)\}$ and corresponding eigenvalues $-A\{\pm \sqrt {\frac{3}{2}},0\}$. The independent time evolution of each eigenstate is then easily followed.

Thus, for initial population in the $s$ state, we have

\begin{subequations}
\begin{eqnarray}
s(t)&=&\frac{1}{3}(1+2\cos[\sqrt {\frac{3}{2}} (A/\omega) \sin \omega t])  \label{subeq1} \\
p(t)&=&\sqrt{\frac{2}{3}} i \sin[\sqrt{\frac{3}{2}}(A/\omega)\sin \omega t] \label{subeq2} \\
d(t)&=&\frac{\sqrt 2}{3}(\cos[\sqrt{\frac{3}{2}}(A/\omega) \sin \omega t]-1). \label{subeq3}
\end{eqnarray}
\end{subequations}
Together with the exponential decrease of the elements of $\eta(t)$, the density matrix can be constructed to reproduce the results in Figs.\ 12-15. It is also clear that when one of the parabolic states is used as a starting point, the density matrix will remain frozen for $\Gamma=0$ and decay monotonically for finite $\Gamma$ as shown in Figs.\ 16 and 17. 

\begin{figure}
\includegraphics[width=3in]{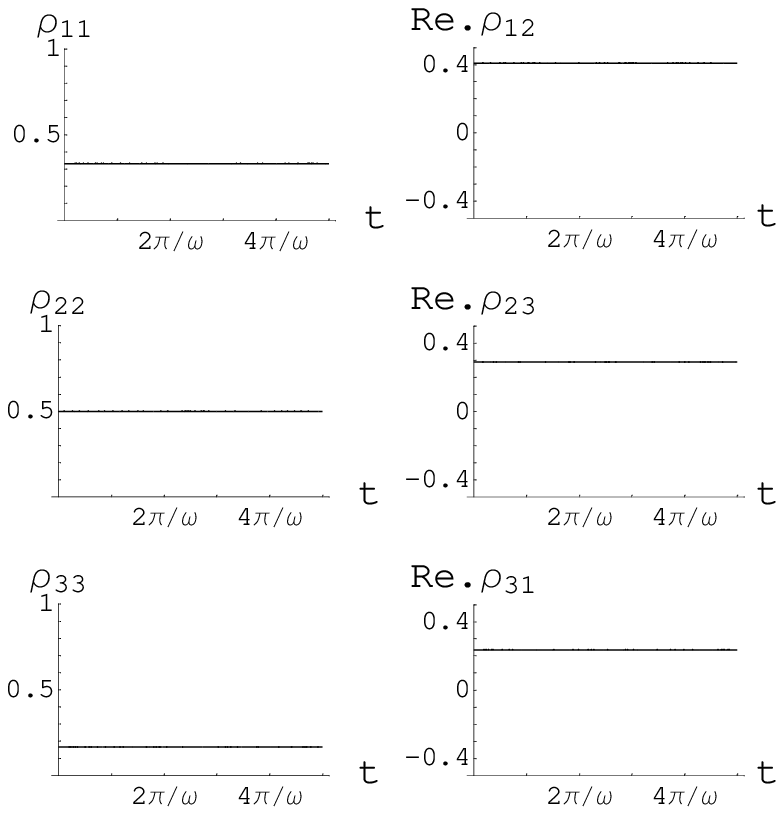}
\caption{Evolution of the $n=3$ states of hydrogen, starting with a Stark eigenstate, and $\Gamma =0$. Such eigenstates remain frozen in their time dependence, in contrast to other initial states as shown in the previous figures.}
\end{figure}

\begin{figure}
\includegraphics[width=3in]{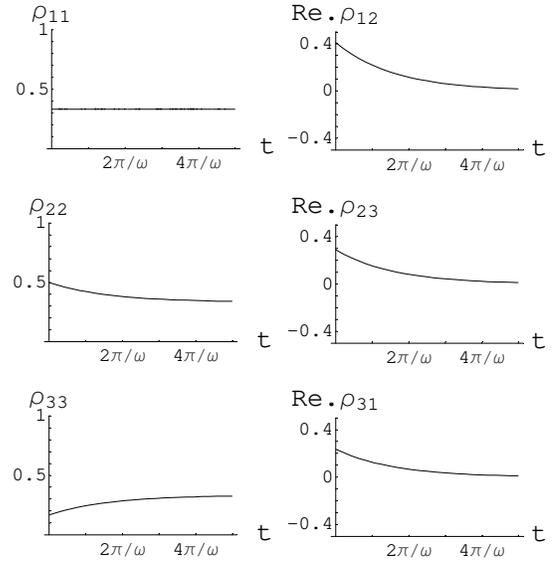}
\caption{same as in Fig.\ 16 except that $\Gamma =0.2$. Note the monotonic evolution from the initial Stark state to the mixed state described by the density matrix $\frac{1}{3}\mathcal{I}$.}
\end{figure} 

A recent paper has presented results similar to the above in Eqs.\ (12) for $n=2, 3$ \cite{ref9}. The results can be readily extended to any $n$ since the expansion of parabolic states in terms of spherical states of hydrogen are well known and given by $3j$-coefficients \cite{ref10}. The occurrence of the ``Floquet form" in Eqs.\ (12), with trigonometric functions whose arguments are themselves a trigonometric function scaled by $A/\omega$, accounts for the appearance of higher frequencies than $\omega$ in Figs.\ 1-8 for stronger driving fields.

We note that such studies of Rydberg atoms in an $n$ manifold under microwave ionization, sometimes with an additional static field, have been of considerable experimental and theoretical interest \cite{ref11, ref12}. 

\section{Summary}

The method of unitary integration as extended to problems involving dissipation and decoherence affords a convenient and powerful way of treating $n$-state systems in time-dependent fields. Through the solution of a single, classical, Riccati equation (first-order in time and quadratically nonlinear), we can follow the evolution of the density matrix in time without any restrictions to infinitesimal steps or time orderings. Results have been presented for three-state systems with examples of coupled pendula and the $n=3$ states of hydrogen in a radiation field.

This work has been supported by the U.S. Department of Energy under Grant No. DE-FG02-02ER46018.

\end{document}